\documentclass[aps,twocolumn,preprintnumbers]{revtex4}
\usepackage{graphicx}

\preprint{The following article has been submitted to Applied Physics Letters}
\begin{document}

\title{Experimental realization of a silicon spin field-effect transistor}

\author{Biqin Huang}
\altaffiliation{bqhuang@udel.edu} \affiliation{Electrical and
Computer Engineering Department, University of Delaware, Newark,
Delaware, 19716}
\author{Douwe J. Monsma}
\affiliation{Cambridge NanoTech Inc., Cambridge MA 02139}
\author{Ian Appelbaum}
\affiliation{ Electrical and Computer Engineering Department,
University of Delaware, Newark, Delaware, 19716}

\begin{abstract}
A longitudinal electric field is used to control the transit time (through an undoped silicon vertical channel) of spin-polarized electrons precessing in a perpendicular magnetic field. Since an applied voltage determines the final spin direction at the spin detector and hence the output collector current, this comprises a spin field-effect transistor. An improved hot-electron spin injector providing $\approx$115\% magnetocurrent, corresponding to at least $\approx$38\% electron current spin polarization after transport through 10$\mu$m undoped single-crystal silicon, is used for maximum current modulation.
\end{abstract}

\maketitle
\newpage

\begin{figure}
  \includegraphics[width=8.5cm,height=4cm]{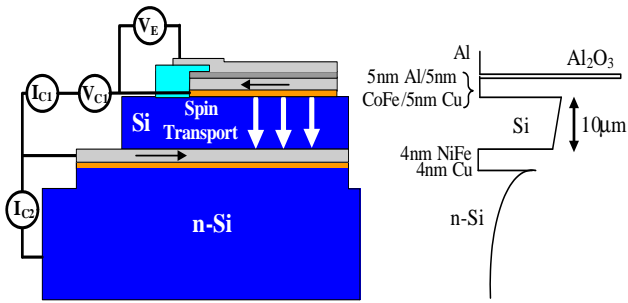}
  \caption{\label{fig:scheme}Schematic illustration of the Si spin field-effect device used in this work (left), and associated conduction band diagram (right). The vertical structure (top to bottom) is 40nm Al/Al$_2$O$_3$/5nm Al/5nm Co$_{84}$Fe$_{16}$/5nm Cu/10 $\mu m$ undoped Si/4nm Ni$_{80}$Fe$_{20}$/4nm Cu/n-Si. Hot electrons are injected by an emitter voltage ($V_E$) from Al ballistically through the Al/Co$_{84}$Fe$_{16}$/Cu anode base and into the conduction band of the 10$\mu m$-thick undoped Si drift layer forming injected current $I_{C1}$. Detection on the other side is with spin-dependent ballistic hot electron transport through the Ni$_{80}$Fe$_{20}$ thin film. Our spin-transport signal is the ballistic current transported into the conduction band of the n-Si collector ($I_{C2}$).}
\end{figure}

The spin field effect transistor (spinFET) proposed by Datta and Das\cite{DATTA1990} has stimulated much research in spin precession-controlled electronic semiconductor devices\cite{ZuticRMP, Awschalom}. Because silicon (Si) has a very long intrinsic electron spin lifetime\cite{Tyryshkin, ZuticPRL} and is the cornerstone of modern semiconductor microelectronics, it could be the materials basis of a future semiconductor spintronics paradigm utilizing these types of devices. However, spintronics techniques which worked so well for other semiconductors, most notably GaAs\cite{Kato2004,Sih2005,Gupta2001,Crowell06, Crowell07}, are ineffective with silicon for both bandstructure and materials growth reasons.\cite{ZuticPRL}

To solve this problem, we have recently demonstrated spin transport in silicon using hot-electron transport through ferromagnetic (FM) metal thin films for all-electrical spin-polarized injection and detection.\cite{appelbaumnature} Because the device design includes rectifying Schottky barriers on either side of the Si transport layer, an applied accelerating voltage induces little spurious current, allowing transit-time control of final spin direction at the spin detector during precession in a perpendicular magnetic field. Two of us have recently proposed to use this effect as the basis of a transit-time spinFET.\cite{appelbaumapl}

To demonstrate the transit-time spinFET, the output collector current magnetocurrent change must be larger than any magnetically-independent current rise induced by accelerating voltage increase. However, successful operation of previously demonstrated devices in this proposed mode is prevented by the low magnetocurrent signal of only $\approx$2\%, and the presence of a small, but significant, rise in collector current with accelerating voltage.

One possible reason for this low spin injection efficiency could be a ``magnetically-dead'' silicide layer\cite{VEUILLEN1987,Tsay19992,Tsay1999,IgorNV} formed between the silicon and ferromagnetic metals used for injector and detector in this device. As we have already demonstrated\cite{bhuang}, by relocating the injector ferromagnetic layer away from the silicon Schottky interface, the spin injection efficiency increased by over an order of magnitude. In this Letter, we demonstrate even higher magnetocurrent in silicon spin transport devices with a further modified injector structure utilizing ballistic spin filtering \cite{Monsma1, Monsma, Jansen} and a Cu interfacial interlayer to prevent silicide formation with the FM layer. We then use this device to realize the transit-time spinFET.

\begin{figure}
    \includegraphics[width=7cm,height=12cm]{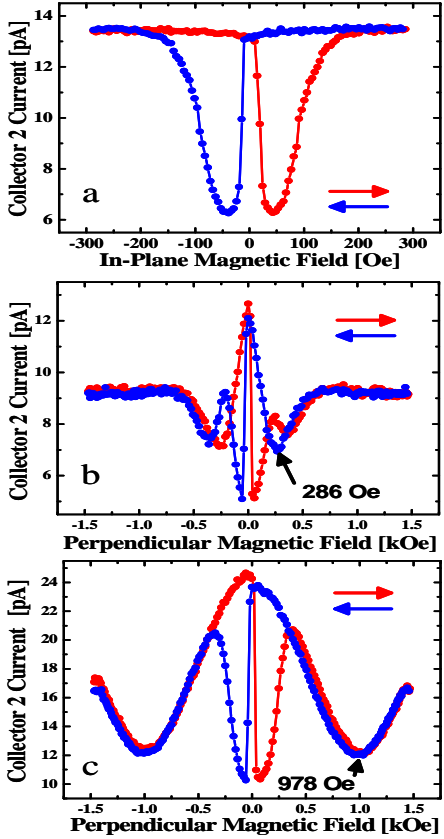}
   \caption{(a) In-plane spin-valve effect for the device with emitter tunnel junction bias $V_E=$-1.6V and $V_{C1}=$0V at 85K, showing $\approx$115\% magnetocurrent ratio. (b) Spin precession and dephasing (Hanle effect) in a perpendicular magnetic field with $V_E=$-1.6V and accelerating voltage $V_{C1}=$0V. (c) Same as in (b), but with $V_{C1}=$3V. }
\end{figure}

A schematic illustration for our improved device in side-view is shown in Fig. 1, together with its associated band-diagram. The injector structure is 40nm Al/Al$_2$O$_3$/5nm Al/5nm Co$_{84}$Fe$_{16}$/5nm Cu. Unpolarized electrons tunneling from the normal metal Al across the Al$_2$O$_3$ oxide barrier are subsequently spin polarized by the hot-electron ballistic spin filtering effect (spin-dependent scattering) through the Co$_{84}$Fe$_{16}$ layer before conduction-band injection over the Cu/Si Schottky barrier.

\begin{figure}
    \includegraphics[width=9cm,height=12cm]{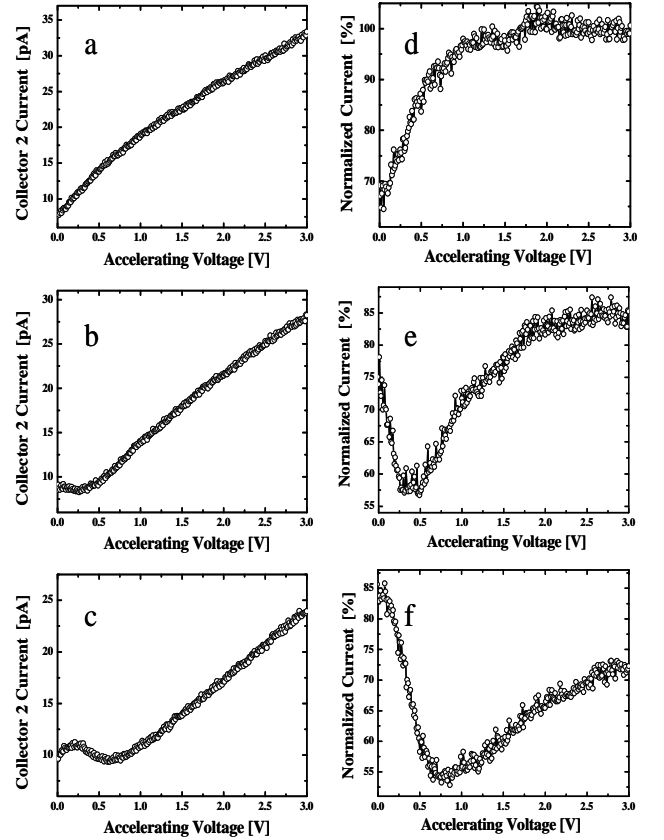}
   \caption{(a)-(c) Spin detection current $I_{C2}$ as a function of accelerating voltage bias $V_{C1}$ in a fixed perpendicular magnetic field. The magnetic field is 191 Oe, 380 Oe and 560 Oe, for (a)-(c), respectively. (d)-(f) shows (a)-(c), respectively, normalized by $I_{C2}$ spectroscopy in zero magnetic field.}
\end{figure}

After vertical transport through the 10 $\mu$m-thick undoped single-crystal silicon device layer, the spin-polarized electrons are ejected from the Si conduction band into the detector FM thin film (Ni$_{80}$Fe$_{20}$) above the Fermi energy. The ballistic component of this hot-electron current is collected by the second Schottky barrier with a n-Si substrate, forming the collector current and spin-transport signal ($I_{C2}$). By manipulating the relative orientation of the injector and detector FM layer magnetizations with an in-plane external magnetic field, $I_{C2}$ can be changed correspondingly. This in-plane spin-valve hysteresis at constant emitter bias $V_E=$-1.6V is shown in Fig. 2(a). The magnetocurrent ratio $MC=(I_{C2}^P-I_{C2}^{AP})/I_{C2}^{AP}$, where the superscripts $P$ and $AP$ refer to parallel and anti-parallel FM injector/detector magnetization configuration, respectively, is approximately 115\%, much higher than in the devices we reported before\cite{appelbaumnature, bhuang}. This magnetocurrent ratio, enabled by (i) avoiding silicide formation with the injector FM, and (ii) using ballistic spin filtering, corresponds to a conduction electron current spin polarization of at least $\mathcal{P}=(I_{C2}^P-I_{C2}^{AP})/(I_{C2}^{P}+I_{C2}^{AP})=MC/(MC+2)\approx 38$\%.

Spin precession measurements of $I_{C2}$ in a perpendicular magnetic field\cite{JOHNSON1988,JOHNSON1985} at different accelerating voltage bias $V_{C1}$ across the Si spin transport layer were performed, as shown in Figs. 2(b) and (c). Due to a small in-plane component of the applied magnetic field, $I_{C2}$ drops when the external perpendicular magnetic field is swept through zero because the Ni$_{80}$Fe$_{20}$ detector magnetization switches at low coercive field. At approximately 500 Oe, the parallel magnetization configuration is regained when the Co$_{84}$Fe$_{16}$ switches. The first extrema away from zero applied field (indicated with arrows) corresponds to magnetic field conditions such that the precession angle $\theta=\omega \tau =\pi$ (so that final spin direction and analyzing FM magnetization are anti-parallel), where $\tau$ is the transit time from injector to detector and $\omega$ is the precession angular frequency $g\mu_B B/\hbar$. In this expression, $g$ is the electron spin g-factor, $\mu_B$ is the Bohr magneton, $B$ is the magnetic field, and $\hbar$ is the reduced Planck constant. Since  $\tau\approx L/(\mu E)$, where $L=10\mu m$ is the transport distance through the undoped Si, $\mu$ is the electron mobility, and $E$ is the electric field, the accelerating voltage controls the transit time and hence final spin precession angle through $E=V_{C1}/L$. As shown in Fig. 2(b) and (c), with an increase of the accelerating voltage bias across the Si spin transport layer from 0V to 3V, the magnetic field corresponding to $\pi$ final spin precession angle increases from 286 Oe to 978 Oe due to the associated reduction of the transit time and the subsequent need for higher precession frequency.

Although Figs. 2(b) and (c) show measurements at fixed electric field under conditions of varying magnetic field, we can alternatively change $\theta$ at fixed perpendicular magnetic field by varying the electric field. Refer to the partial parallel magnetization curve in the Hanle measurements, which corresponds to the right-left (blue) sweep in positive field shown in Figs. 2(b) and (c). From these measurements, it can be seen that if the fixed perpendicular magnetic field is smaller than 286 Oe, an increase of $V_{C1}$ past 0V causes a continual increase of $I_{C2}$ because the precession angle does not pass through $\pi$. For fixed perpendicular magnetic fields slightly larger than 286 Oe, $I_{C2}$ will first decrease with increased $V_{C1}$ as the precession angle approaches $\pi$, and then continually increase.

This electric field dependence of $I_{C2}$ at fixed perpendicular magnetic fields, and in a parallel injector/detector magnetization configuration, is shown in Figs. 3(a)-(c) for 191 Oe, 380 Oe and 560 Oe, respectively. Although there is an initial decrease in the measured current at applied fields above 286 Oe as predicted, an ascending trend is dominant due to the increase of injected current ($I_{C1}$) which drives $I_{C2}$.\cite{SiSpinJAP} This is likely due to enhanced hot-electron collection efficiency under applied bias.\cite{SZE}

One straightforward solution to this problem is to continue to improve the spin injection efficiency and output current magnitude so that the $I_{C2}$ change due to precession angle control will make the increase caused by $V_{C1}$ dependence negligible. However, we can eliminate this effect artificially by normalizing Figs 3(a)-(c) with $I_{C2}(V_{C1})$ in zero magnetic field. The result, shown in Figs. 3(d)-(f), respectively, agrees very well with our expectation based on the analysis of spin precession measurements (Figs. 2(b) and (c)).

In summary, we have presented measurements of a silicon spin transport device showing output current modulation through voltage control of spin precession. Therefore, it comprises successful operation as a transit-time spinFET. This was enabled by an improved spin-polarized hot-electron injector utilizing ballistic spin filtering. Our work presents dual ways to manipulate the spin direction in spintronic devices: magnetic, through precession frequency $\omega$, and electric, through transit time $\tau$.

This work was supported in part by DARPA/MTO.


\begin{thebibliography}{10}

\bibitem{DATTA1990}
S. Datta and B. Das, Appl. Phys. Lett. {\bf{56}}, 665 (1990).

\bibitem{ZuticRMP}
I. \v{Z}uti\'c, J. Fabian and S. Das Sarma, Rev. Mod. Phys. {\bf{76}}, 323 (2004).

\bibitem{Awschalom}
D. D. Awschalom and M. E. Flatt\'e, Nature Phys. {\bf{3}}, 153 (2007).

\bibitem{Tyryshkin}
A. M. Tyryshkin, S. A. Lyon, A. V. Astashkin, and A. M. Raitsimring, Phys. Rev. B {\bf{68}}, 193207 (2003).

\bibitem{ZuticPRL}
I. \v{Z}uti\'c, J. Fabian, and S. C. Erwin, Phys. Rev. Lett. {\bf{97}}, 026602 (2006)

\bibitem{Kato2004}
Y. K. Kato, R.C. Myers, A.C. Gossard, and D. D. Awschalom, Science
{\bf{306}}, 1910 (2004).

\bibitem{Sih2005}
V. Sih, R. C. Myers, Y. K. Kato, W. H. Lau, A. C. Gossard, and D. D.
Awschalom, Nature Phys. {\bf{1}}, 31 (2005).

\bibitem{Gupta2001}
J. A. Gupta, R. Knobel, N. Samarth, and D. D. Awschalom, Science
{\bf{292}}, 2458 (2001).

\bibitem{Crowell06}
X. Lou, C. Adelmann, M. Furis, S. A. Crooker, C. J. Palmstrom, and P. A. Crowell, Phys. Rev. Lett. {\bf{96}}, 176603 (2006).

\bibitem{Crowell07}
X. Lou, C. Adelmann, S. A. Crooker, E. S. Garlid, J. Zhang, S. M. Reddy, S. D. Flexner, C. J. Palmstrom, and P. A. Crowell, Nature Phys. {\bf{3}}, 197 (2007).


\bibitem{appelbaumnature}
I. Appelbaum, B. Huang, and D. Monsma, Nature {\bf{447}}, 295 (2007).

\bibitem{appelbaumapl}
I. Appelbaum and D. Monsma, Transit-Time Spin
Field-Effect-Transistor (2007), URL
http://arxiv.org/abs/cond-mat/0703300

\bibitem{VEUILLEN1987}
J. Y. Veuillen, J. Derrien, P. A. Badoz, E. Rosencher, and C.
Danterroches, Appl. Phys. Lett. {\bf{51}}, 1448 (1987).

\bibitem{Tsay19992}
J. S. Tsay and Y. D. Yao, Appl. Phys. Lett. {\bf{74}}, 1311 (1999).

\bibitem{Tsay1999}
J. S. Tsay, C. S. Yang, Y. Liou, and Y. D. Yao, J. Appl. Phys.
{\bf{85}}, 4967 (1999).

\bibitem{IgorNV}
I. \v{Z}uti\'c and J. Fabian, Nature {\bf{447}}, 269 (2007).

\bibitem{bhuang}
B. Huang, L. Zhao, D. Monsma and I. Appelbaum, 35\% magnetocurrent
with spin transport through Si (2007), URL
http://arxiv.org/abs/0704.3949

\bibitem{Monsma1}
D.J. Monsma, J.C. Lodder, Th. J. A. Popma, B. Dieny, Phys.
Rev. Lett. {\bf{74}}, 5260 (1995).

\bibitem{Monsma}
D.J. Monsma, R. Vlutters, and J.C. Lodder, Science {\bf{281}}, 407
(1998).

\bibitem{Jansen}
R. Jansen, J. Phys. D {\bf{36}}, R289 (2003).

\bibitem{JOHNSON1988}
M. Johnson and R. H. Silsbee, Phys. Rev. B  {\bf{37}}, 5326 (1988).

\bibitem{JOHNSON1985}
M. Johnson and R. H. Silsbee, Phys. Rev. Lett. {\bf{55}}, 1790 (1985).

\bibitem{SiSpinJAP}
B. Huang, D. Monsma and I. Appelbaum, Spin lifetime in silicon in the presence of parasitic electronic effects (2007), URL http://arxiv.org/abs/0704.3928

\bibitem{SZE}
C. Crowell and S. Sze, in {\it{Physics of thin films}}, edited by G. Hass and R. Thun (Academic press, 1967), vol. 4.

\end{thebibliography}
\end{document}